\documentclass[10pt,
               twocolumn,
               showpacs,
               superscriptaddress,
               floatfix,
               longbibliography,
               aps,
               prl]{revtex4-2}

\usepackage[english]{babel}

\usepackage{graphicx,wrapfig}
\graphicspath{{./}}

\usepackage{amsmath,amssymb,bm,physics,upgreek,cancel,extarrows}
\usepackage{booktabs}

\usepackage[colorlinks=true,
            allcolors=blue]{hyperref}

\usepackage{xr}

\usepackage{acro}
\DeclareAcronym{sm}{
    short=SM,
    long=Supplemental Material,
}
\DeclareAcronym{em}{
    short=EM,
    long=End Matter,
}

\usepackage{comment}
\usepackage{subfiles}
\usepackage{lipsum}

\newcommand{\pii}{\uppi}
\newcommand{\rc}{r_\mathrm{c}}
\newcommand{\Calcium}{\mathrm{^{40}Ca^+}}
\newcommand{\Strontium}{\mathrm{^{88}Sr^+}}
\newcommand{\Rubidium}{\mathrm{^{85}Rb}}

\begin{document}

\title{Anomalously enhanced lifetimes of low angular momentum Rydberg states in singly charged alkaline-earth metal ions}

\author{Simon Euchner}
\affiliation{Institut f\"ur Theoretische Physik and Center for Integrated Quantum Science and Technology, Universit\"at Tübingen, Auf der Morgenstelle 14, 72076 T\"ubingen, Germany}
\author{Weibin Li}
\affiliation{School of Physics and Astronomy and Centre for the Mathematics and Theoretical Physics of Quantum Non-Equilibrium Systems, The University of Nottingham, Nottingham, NG7 2RD, United Kingdom}
\author{Igor Lesanovsky}
\affiliation{Institut f\"ur Theoretische Physik and Center for Integrated Quantum Science and Technology, Universit\"at Tübingen, Auf der Morgenstelle 14, 72076 T\"ubingen, Germany}
\affiliation{School of Physics and Astronomy and Centre for the Mathematics and Theoretical Physics of Quantum Non-Equilibrium Systems, The University of Nottingham, Nottingham, NG7 2RD, United Kingdom}

\begin{abstract}
    Trapped ions excited to high-lying electronic states, so-called Rydberg states, open new opportunities for quantum simulation and quantum computing. Generally, the fidelity of quantum coherent operations critically depends on the longevity of Rydberg states. However, scaling laws predict that the lifetimes of Rydberg states in singly charged alkaline-earth metal ions are 16 times shorter, compared to their neutral atom counterparts. Here, we show that this is not generally the case. We report an anomalous lifetime enhancement of certain low angular momentum ionic Rydberg series by factors larger than eight. The anomaly is present at both zero and finite temperature, although it is caused by different mechanisms. At zero temperature, the anomalously enhanced lifetimes are caused by accidental cancellations of the relevant dipole transition matrix elements, while at room temperature the anomaly originates from the enlarged energetic separation of ionic Rydberg levels with respect to neutral-atom levels.
\end{abstract}

\maketitle


\textit{Introduction.} Neutral alkali-metal atoms excited to high-lying electronic states, so-called \emph{Rydberg} states \cite{gallagher1988}, constitute one of today's most advanced platforms for controlling single quantum mechanical degrees of freedom. The key ingredients responsible for the success of Rydberg atoms are their long lifetimes together with strong electronic state-dependent dipolar interactions \cite{saffman2010,sibalic2018,browaeys2020}. Rydberg states can also be excited in trapped ion crystals \cite{mueller2008,schmidtkaler2011,mallweger2025}. Compared to arrays of neutral atoms, crystals of Rydberg ions may offer certain advantageous features: they provide strong, electronic-state-independent confinement \cite{pokorny2019,pokorny2020} for both low-lying states and Rydberg states and they offer collective vibrational modes which can be spatially delocalised over the whole ion crystal. This opens new opportunities for the quantum simulation of so-called \emph{conical intersections} in molecular potential-energy surfaces \cite{gambetta2021,chaudhary2024,belfakir2026A,belfakir2026B,belfakir2026C}, the study of excitation energy transport \cite{euchner2026}, the creation of electron-phonon entanglement \cite{wilkinson2024}, the exploration of electronic state-dependent molecular conformational changes \cite{mallweger2025}, and the realisation of sub-microsecond entangling gates \cite{li2014,zhang2020,wilkinson2025} for digital quantum computing. Moreover, the possibility to cool the ion crystal's vibrational degrees of freedom \cite{mao2021,so2024,so2026}, in principle even in the Rydberg state, may open a route towards the study of long-time dissipative dynamics.

A potential downside of quantum simulation and computation with ionic Rydberg states is that their lifetimes are generally expected to be substantially shorter than those of Rydberg states in neutral atoms. This is seen as follows: the lifetimes of Rydberg states scale as $\tau {\sim} \mu^{-2}_\mathrm{i,f}\lambda^3_\mathrm{i,f}$ \cite{friedrich2017}. Here $\mu^2_\mathrm{i,f}$ is the radial transition dipole matrix element \cite{sibalic2017} between the initial Rydberg state $\ket*{i}$ and the final state $\ket*{\mathrm{f}}$, and $\lambda_\mathrm{i,f}$ is the wavelength of the photon emitted upon decay. Rydberg states in alkali-metal atoms and alkaline-earth metal ions closely resemble Rydberg states of Hydrogenic ions with effective nuclear charge $Z_\mathrm{eff}=1$ and $Z_\mathrm{eff}=2$, respectively. The analytical solution for the eigenstates of the Hydrogenic ion with nuclear charge $Z_\mathrm{eff}$ leads to the so-called \emph{Hydrogenic} scaling laws $\mu_\mathrm{i,f}{\sim}Z^{-1}_\mathrm{eff}$ and $\lambda_\mathrm{i,f}{\sim}Z^{-2}_\mathrm{eff}$ \cite{friedrich2017,higgins2019}. Therefore, the lifetimes of Rydberg states are expected to follow the Hydrogenic scaling law \cite{higgins2019}
\begin{equation}\label{eq:LifetimeScalingLawZeff}
    \tau(Z_\mathrm{eff}) \propto Z^{-4}_\mathrm{eff}
\,,
\end{equation}
which predicts that the lifetime of Rydberg states in singly charged ions is reduced by a factor of 16 compared to the lifetime of Rydberg states in neutral atoms.

In this work, we show that certain low angular momentum Rydberg series in alkaline-earth metal ions do not follow the scaling law \eqref{eq:LifetimeScalingLawZeff}, but are significantly larger than expected, even reaching the lifetime values of Rydberg states of neutral atoms. Therefore, we refer to lifetimes that do not follow \eqref{eq:LifetimeScalingLawZeff} as \emph{anomalously enhanced}. As an illustration of this enhancement we systematically compare the lifetimes of $\Rubidium$ and $\Strontium$ and show that at zero temperature Rydberg P states are anomalously long-lived. This anomaly is caused by an accidental cancellation in the transition matrix elements of the relevant decay pathways. Interestingly, this anomalous longevity persists even at room temperature. Here, the anomaly is underpinned by the enlarged energetic separation of ionic Rydberg levels compared to those of neutral atoms.

\begin{figure*}[t]
    \centering
    \includegraphics{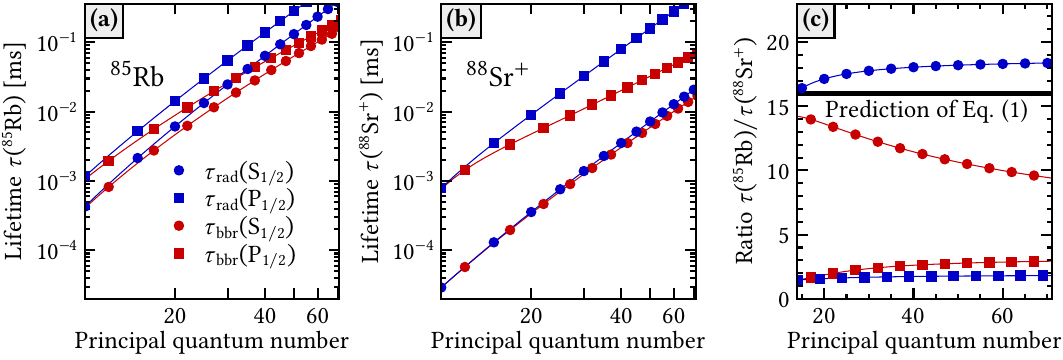}
    \caption{\textbf{Rydberg-state lifetimes.} (a) Numerically computed lifetimes of Rydberg S (dots) and P (squares) states of $\mathrm{^{85}Rb}$. Blue and red data show the radiative ($T=0\,\mathrm{K}$) lifetimes \eqref{eq:RadiativeLifetime}, $\tau_\mathrm{rad}$, and the room-temperature ($T=300\,\mathrm{K}$) lifetimes \eqref{eq:ReducedLifetime}, $\tau_\mathrm{bbr}$, respectively. The data points are connected to guide the eye. (b) Analogous plot to panel (a) but for $\mathrm{^{88}Sr^+}$. (c) Comparison of lifetimes. The panel shows the ratio $\tau(\Rubidium)\slash\tau(\Strontium)$ between the lifetimes $\tau(\Rubidium)$ and $\tau(\Strontium)$ of Rydberg states in $\Rubidium$ and $\Strontium$. The Hydrogenic scaling law \eqref{eq:LifetimeScalingLawZeff} predicts the constant ratio of 16, indicated by the black horizontal line. The lifetimes shown in panels (a) and (b) are tabulated in the \ac{sm} \cite{supmat} and can be compared to numerical calculations of Rydberg-state lifetimes for neutral Group-I and singly charged Group-II elements conducted in Refs.~\cite{theodosiou1984,he1990,beterov2009,beterov2009erratum,glukhov2013,glukhov2016}. Benchmarks of the numerically computed lifetimes for $\Rubidium$ against experimental data can be found in the \ac{em}.}
    \label{fig:LifetimesRbSr}
\end{figure*}


\textit{Calculation of Rydberg state lifetimes.} In this section, we describe how we numerically compute the lifetimes of Rydberg states in alkali-metal atoms (Group I) and alkaline-earth metal ions (Group II). For concreteness, we focus on $\Rubidium$ and $\Strontium$, both employed in experiments with Rydberg atoms and ions \cite{bendkowsky2009,higgins2019}. These two species are part of the same isoelectronic sequence $\mathrm{[Kr]5s^1}$, such that their electronic properties can be compared directly.

To numerically compute the lifetimes of the relevant Rydberg levels, we start by computing the eigenenergies and eigenstates for both $\Rubidium$ and $\Strontium$. This is done by diagonalising the following single-electron model Hamiltonian that describes the dynamics of the valence electron:
\begin{equation}\label{eq:Hamiltonian}
    H_\mathrm{val}
    =
    \frac{\bm{p}^2}{2m}
    +
    V_s(r)
\,.
\end{equation}
The first term is the kinetic energy of the valence electron with mass $m$. The second term is a so-called \emph{parametric model potential} $V_s$, which depends on the atom or ion species $s=\mathrm{^{85}Rb},\mathrm{^{88}Sr^+}$. This potential accounts for the fact that the nucleus is `screened' by the ground-state electron cloud. The size of the electron cloud is characterised by the cutoff radius $\rc$ \cite{marinescu1994,aymar1996}. At distances $r\gg\rc$, the valence electron experiences the Coulomb potential $V_s(r\gg\rc)\approx-Z_\mathrm{eff}e^2\slash(4\pii\varepsilon_0r)$. Inside the effective nucleus, where $r\lesssim\rc$, $V_s$ includes short-ranged corrections. These corrections account for the electric polarisability of the electron cloud, its finite-size characterised by $\rc$, and the gradual transition between a Coulomb potential associated with the nuclear charge $Z$ and the effective charge $Z_\mathrm{eff}$, as $r$ increases far beyond the cutoff radius $\rc$ \cite{marinescu1994,aymar1996,friedrich2017}. In this work we employ the parametric model potentials $V_s$ from Refs.~\cite{marinescu1994,aymar1996} together with the spin-orbit interaction used in Ref.~\cite{wilkinson2025}, which follows from relativistic perturbation theory \cite{condon1951}. To numerically diagonalise the Hamiltonian we use the software \emph{AlkCalc} \cite{alkcalc}.

With the numerical results for the eigenenergies and the eigenstates we compute the lifetimes as follows: at zero temperature and within the dipole approximation, the inverse radiative lifetime $\tau^{-1}_\mathrm{rad}$ of the initial Rydberg state $\ket*{\mathrm{i}}$ is given by the inverse sum of Einstein-A coefficients \cite{friedrich2017}:
\begin{equation}\label{eq:RadiativeLifetime}
    \tau^{-1}_\mathrm{rad}
    =
    \sum_{\mathrm{f}}
    A_\mathrm{i\to{f}}
\,, \ \ \
    \frac{\hbar}{E_\mathrm{H}}
    A_\mathrm{i\to{f}}
    =
    \frac{4}{3} (2\pii)^3
    F_{l_\mathrm{i},l_\mathrm{f}}
    \frac{a_\mathrm{B}\mu^2_\mathrm{i,f}}
         {\lambda^3_{\mathrm{i,f}}}
\,.
\end{equation}
The Einstein-A coefficient $A_\mathrm{i\to{f}}$ \cite{einstein1917} is the rate at which the initial state $\ket*{\mathrm{i}}$ at energy $E_\mathrm{i}$ decays to the final electronic state $\ket*{\mathrm{f}}$ at energy $E_\mathrm{f}<E_\mathrm{i}$. Further, $\alpha\approx1\slash{137}$ is the fine-structure constant, $a_\mathrm{B}\approx53\,\mathrm{pm}$ is Bohr's radius, $\lambda_\mathrm{i,f}=hc\slash\abs*{E_\mathrm{f}-E_\mathrm{i}}$ is the wavelength of the photon emitted upon decay, $\hbar=h\slash(2\pii)$ is the reduced Planck constant, and $E_\mathrm{H}$ is the Hartree energy unit \cite{friedrich2017}. The factor $F_{l_\mathrm{i},l_\mathrm{f}}$ stems from the angular part of the dipole transitions matrix element and depends on the orbital angular momentum quantum numbers $l_\mathrm{i}$ and $l_\mathrm{f}$ of the initial and the final states. It can be calculated using SU(2) angular momentum algebra relations stated in Ref.~\cite{varshalovic1988}. Finally, $\mu_\mathrm{i,f}$ in Eq.~\eqref{eq:RadiativeLifetime} is the radial dipole transition matrix element \cite{sibalic2017} between the initial and the final state.

In a room-temperature (black-body) environment ($T=300\,\mathrm{K}$) the average number of photons with wavelength $\lambda$ at temperature $T$ is $N_\mathrm{ph}(\lambda,T)=[\exp(hc\slash[k_\mathrm{B}T\lambda])-1]^{-1}$. These photons cause radiative transitions from the initial Rydberg state $\ket*{\mathrm{i}}$ to other Rydberg states $\ket*{\mathrm{f}}$, both lower ($E_\mathrm{f}<E_\mathrm{i}$) and higher ($E_\mathrm{f}>E_\mathrm{i}$) in energy. Such transitions depopulate the initial Rydberg state and thereby reduce the radiative lifetime \eqref{eq:RadiativeLifetime}. The total lifetime $\tau_\mathrm{bbr}$, reduced by black-body radiation, is calculated via \cite{friedrich2017}
\begin{equation}\label{eq:ReducedLifetime}
    \tau^{-1}_\mathrm{bbr}
    =
    \tau^{-1}_\mathrm{rad}
    +
    \sum_{\mathrm{f}}
    N_\mathrm{ph}(\lambda_\mathrm{i,f},T)
    A_{\mathrm{i\to{f}}}
\,,
\end{equation}
where the sum is over all possible final states $\ket*{\mathrm{i}}$, both higher and lower in energy than $\ket*{\mathrm{i}}$.

To benchmark the numerically computed lifetimes we compare them in the \ac{em} to experimental results for $\Rubidium$ at room temperature. Since the results agree well with the experimental data, we have confidence in our numerical method \cite{alkcalc}.

\begin{figure}[t]
    \centering
    \includegraphics{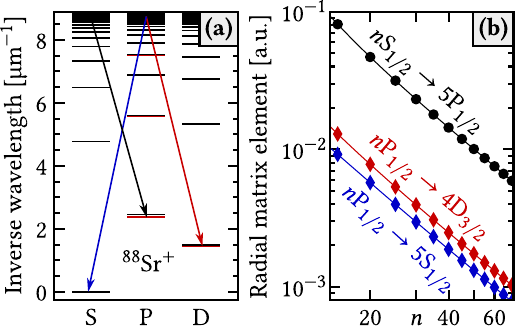}
    \caption{\textbf{Electronic spectrum and transition matrix elements of $\bm{\Strontium}$.} (a) Electronic bound-state spectrum. Red levels show the energy for $j=\abs*{l-1\slash{2}}$ and black levels for $j=l+1\slash{2}$. Arrows indicate the dominant radiative decay channels for the Rydberg S and P manifolds. (b) Radial transition dipole matrix elements \cite{sibalic2017} $\mu_\mathrm{i,f}$ [cf. Eq.~\eqref{eq:RadiativeLifetime}] as a function of the principal quantum number. For each decay channel in panel (a) the radial dipole matrix elements are shown. The colours indicate to which decay channel (arrows) in panel (a) the matrix elements correspond to. Dots and diamonds correspond to transitions out of the Rydberg S and P manifolds, respectively. The data points are connected to guide the eye.
    }
    \label{fig:Mechanism}
\end{figure}


\textit{Anomalously enhanced lifetimes.} In the following we show that certain low angular momentum Rydberg series have anomalously enhanced lifetimes, for which the Hydrogenic scaling law \eqref{eq:LifetimeScalingLawZeff} breaks down. For concreteness, we focus on $\Strontium$, which is currently employed in experiments with Rydberg ions \cite{higgins2019}. However, in the \ac{sm} \cite{supmat} we tabulate lifetimes and give formulas to calculate them for $\Strontium$, $\Calcium$, and $\Rubidium$. These results show that also for $\Calcium$, an isotope employed in Ref.~\cite{schmidtkaler2011}, Rydberg P states can have lifetimes up to ${\sim}20$ times as high as S states with the same principal quantum number. For simplicity, we focus on the experimentally relevant \cite{higgins2019,schmidtkaler2011} Rydberg states $n\mathrm{S}_{1\slash2}$ and $n\mathrm{P}_{1\slash2}$. The discussion is split into two parts: first, we discuss the radiative lifetimes at zero-temperature, relevant for experiments in cryogenic environments \cite{pagano2018,zhang2025}. Subsequently, we focus on lifetimes in a room-temperature environment.

At zero temperature, we find that Rydberg P states have anomalously enhanced radiative lifetimes. This can be seen in the following way: the lifetimes for $\Rubidium$ in Fig.~\ref{fig:LifetimesRbSr}(a) [blue data] show that Rydberg S and P states have similar lifetimes. In contrast, the lifetimes for $\Strontium$ in Fig.~\ref{fig:LifetimesRbSr}(b) [blue data] show that S states (blue dots) have much shorter lifetimes than P states (blue squares). To test the scaling law \eqref{eq:LifetimeScalingLawZeff} we plot the ratios $\tau(\Rubidium)\slash\tau(\Strontium)$ of the lifetimes in Fig.~\ref{fig:LifetimesRbSr}(c) [blue data]. The numerical data shows that Rydberg S states approximately follow the scaling law \eqref{eq:LifetimeScalingLawZeff}, while ionic Rydberg P states have much longer lifetimes than predicted. This is indicated by the fact that the ratios for S states (blue dots) lie close to the predicated constant value of 16, whereas for P states (blue squares) the ratio is much closer to unity. Therefore, we refer to the ionic Rydberg P-state lifetimes as \emph{anomalously enhanced}.

In a room temperature environment at $T=300\,\mathrm{K}$, we find that both Rydberg S and P states have anomalously enhanced lifetimes. To illustrate this we show in Figs.~\ref{fig:LifetimesRbSr}(a) and (b), analogously to the zero-temperature case, the lifetimes of Rydberg S (red dots) and Rydberg P (red squares) states in $\Rubidium$ and $\Strontium$. Computing the ratio $\tau(\Rubidium)\slash\tau(\Strontium)$ [red data in Fig.~\ref{fig:LifetimesRbSr}(c)] indeed shows that the lifetimes of both Rydberg S \emph{and} Rydberg P states is anomalously enhanced, and the enhancement of the S-state lifetime even grows with the principal quantum number. The latter is evident by the fact that the ratios of the S-state lifetimes (red dots) decrease far below the prediction of Eq.~\eqref{eq:LifetimeScalingLawZeff}. In the remainder of this work, we shed light on the enhancement mechanisms underlying the anomalous lifetime enhancement at zero and room temperature.


\textit{Enhancement mechanism at zero temperature.} At zero temperature, the anomalously enhanced lifetimes of ionic Rydberg P states occurs due to accidental cancellations of the dipole transition matrix elements relevant for spontaneous emission. To rationalise this, we recall that the lifetime is given by the inverse sum over the Einstein-A coefficients $A_\mathrm{i\to{f}}$ --- the decay rates from the Rydberg state $\ket*{\mathrm{i}}$ to the final states $\ket*{\mathrm{f}}$. These coefficients scale as $A_\mathrm{i\to{f}}{\sim}\mu^2_\mathrm{i,f}\lambda^{-3}_\mathrm{i,f}$ \cite{sibalic2017,friedrich2017} with the radial transition matrix element $\mu_\mathrm{i,f}$ \cite{sibalic2017} and the wavelength $\lambda_\mathrm{i,f}$ of the photon emitted upon decay. A consequence of the scaling with the wavelength $\lambda_\mathrm{i,f}$ is that the dominant decay channels for spontaneous emission are transitions to low-lying electronic states \cite{sibalic2017}. In the bound-state energy spectrum of $\Strontium$ in Fig.~\ref{fig:Mechanism}(a) the dominant channels for the relevant Rydberg S and P states are indicated by the black, blue, and red arrows. The radial transition matrix elements in panel (b), associated with the three decay channels, show that the transition matrix elements for decay of P states are indeed approximately an order of magnitude smaller compared to those of S states. The small matrix elements $\mu_\mathrm{i,f}$ suppress the Einstein-A coefficients $A_\mathrm{i\to{f}}{\sim}\mu^2_\mathrm{i,f}$ and thus lead to the anomalously enhanced lifetimes of Rydberg P states shown in Figs.~\ref{fig:LifetimesRbSr}(b) and (c).

Interestingly, the scaling of the Einstein-A coefficients $A_\mathrm{i\to{f}}{\sim}\lambda^{-3}_\mathrm{i,f}$ decreases the Rydberg P-state lifetime compared to that of S states. This can be seen by the fact that the inverse wavelengths $\lambda^{-1}_\mathrm{i,f}$ associated with the dominant decay channels for Rydberg S states (black arrow) are smaller compared to the inverse wavelengths of the relevant channels for P states  (blue and red arrows). However, the numerical results in Fig.~\ref{fig:LifetimesRbSr}(b) show that ionic Rydberg P states have longer lifetimes than S states. This shows that the accidental cancellations in the matrix elements not only compensate this effect, but even enhance the lifetimes of ionic Rydberg P states beyond those of Rydberg S states.

We remark that the accidental cancellation in the matrix elements shown in Fig.~\ref{fig:Mechanism}(b) aligns well with experimental observations reported in the literature: Lange \emph{et al.} \cite{lange1991} observed near-zero oscillator strengths for the transitions to Rydberg P states in singly charged Strontium. Further, Jones and Gallagher \cite{jones1989} observed the same phenomenon in singly charged Barium, which they could attribute to a zero in the radial transition dipole matrix element. Finally, we remark that the small matrix elements are not a \emph{Cooper minimum} \cite{beterov2012} --- an accidental cancellation of the transition matrix element that only appears for certain principal quantum numbers. The anomaly shown by our results is stronger: the cancellation is consistently present across the \emph{whole} range of considered principal quantum numbers.

\begin{figure}[t]
    \centering
    \includegraphics{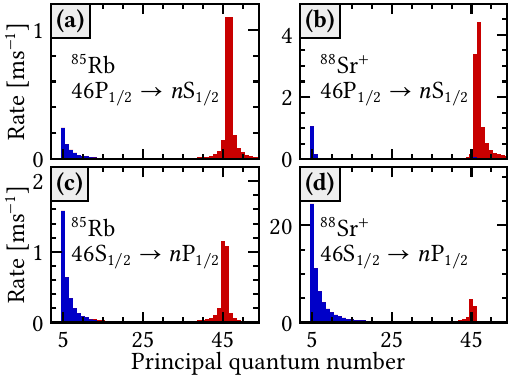}
    \caption{\textbf{Rydberg-state depopulation rates.} (a) Depopulation rates of the Rydberg state $46\mathrm{P}_{1\slash{2}}$ to the states $n\mathrm{S}_{1\slash{2}}$ in $\Rubidium$. The decay rate for spontaneous emission, $A_\mathrm{i\to{f}}$, is shown in blue. The depopulation rates $N_\mathrm{ph}(\lambda_\mathrm{i,f},T)A_\mathrm{i\to{f}}$ associated with transitions induced by black-body radiation are shown in red, where $N_\mathrm{ph}(\lambda_\mathrm{i,f},T)$ is the average number of photons with wavelength $\lambda_\mathrm{i,f}$ at temperature $T$. (b) Same as in (a) but for $\Strontium$. (c) Same as in (a) but for transitions depopulating the Rydberg S state. (d) Same as in (c) but for $\Strontium$.
    }
    \label{fig:Rates}
\end{figure}


\textit{Enhancement mechanism at room temperature.} In a room-temperature environment the ratios in Fig.~\ref{fig:LifetimesRbSr}(c) show that the lifetimes of both Rydberg S and P states are anomalously enhanced, i.e., do not follow the Hydrogenic scaling law \eqref{eq:LifetimeScalingLawZeff}. This is caused by an increased energetic separation of ionic Rydberg energy levels compared to neutral atoms. To shed light on this, we first focus on Rydberg P states and show their depopulation rates for $\Rubidium$ and $\Strontium$ in Fig.~\ref{fig:Rates}(a) and (b), respectively. The room-temperature lifetime of Rydberg P states is dominated by transitions induced by black-body radiation. This can be seen by the fact that the depopulation rates $N_\mathrm{ph}(\lambda_\mathrm{i,f},T)A_\mathrm{i\to{f}}$ associated with transitions induced by black-body radiation (red bars), dominate the spontaneous emission rates $A_\mathrm{i\to{f}}$ (blue bars). Therefore, according to Eq.~\eqref{eq:ReducedLifetime}, the total lifetime of Rydberg P states can be approximated as
\begin{equation}\label{eq:ApproximateLifetimesOfPStates}
    \tau^{-1}_\mathrm{bbr}
    \approx
    \sum_{\mathrm{f}}
    N_\mathrm{ph}(\lambda_\mathrm{i,f},T)A_\mathrm{i\to{f}}
    \propto
    N_\mathrm{ph}(\lambda_\mathrm{i,f},T) Z^{4}_\mathrm{eff}
\,.
\end{equation}
At the wavelengths $\lambda_\mathrm{i,f}$ relevant for transitions between Rydberg states, the average number of photons is well approximated as $N_\mathrm{ph}(\lambda_\mathrm{i,f},T)\approx\lambda_\mathrm{i,f}k_\mathrm{B}T\slash{(hc)}\propto{Z}^{-2}_\mathrm{eff}$, where $k_\mathrm{B}$ is Boltzmann's constant, $c$ is the speed of light, and $h$ is Planck's constant. Inserting this result into Eq.~\eqref{eq:ApproximateLifetimesOfPStates} yields the following scaling law for lifetimes dominated by black-body radiation:
\begin{equation}\label{eq:BlackBodyScalingLaw}
    \tau_\mathrm{bbr}
    \propto
    Z^{-2}_\mathrm{eff}
\,.
\end{equation}
This predicts that for Rydberg P states $\tau(\Rubidium)\slash\tau(\Strontium)\approx4$, which is consistent with the data in Fig.~\ref{fig:LifetimesRbSr}(c) [red squares], showing a ratio of approximately 3.

Finally, we consider Rydberg S states, whose rates for spontaneous emission and depopulation through black-body radiation are exemplarily shown for S-P transitions in Figs.~\ref{fig:Rates}(c) and (d). In contrast to the P states, the lifetimes of the S states are not dominated by transitions to nearby Rydberg states due to black-body radiation and the scaling law \eqref{eq:BlackBodyScalingLaw}. Nevertheless, there is a weak enhancement over the Hydrogenic scaling law \eqref{eq:LifetimeScalingLawZeff}.


\textit{Summary and outlook.} Certain low angular momentum Rydberg series in singly charged alkaline-earth metal ions have unexpectedly long lifetimes, which can even become comparable to those of neutral atoms. Here, we have shed light on the mechanisms underlying this anomalous lifetime enhancement in both zero- and room-temperature environments. An important question is how this longevity can be exploited to enhance quantum computation and simulation protocols. For this it will be important to understand how much of this property persists in the presence of strong electromagnetic fields, since ions are typically confined in Paul \cite{paul1990} or Penning \cite{vogel2024} traps. The latter feature magnetic flux densities on the order of few Tesla, which drastically alters the electronic bound-state structure of the trapped Rydberg ions \cite{knudsen1984,wunner1986,wunner1987,pohl2009,martins2026}.


\textit{Acknowledgements.} S.E. thanks W. S. Martins and J. W. P. Wilkinson for fruitful discussions on Rydberg ions. This work was supported by the European Union through the ERC grant OPEN-2QS (Grant No. 101164443) and the Deutsche Forschungsgemeinschaft within the research unit FOR 5413 (Grant No. 465199066). We also acknowledge funding through JST-DFG 2024: Japanese-German Joint Call for Proposals on “Quantum Technologies” (Japan-JST-DFG-ASPIRE 2024) under JST Grand No. JPMJAP24C2 and DFG Grant No. 554561799. W.L. acknowledges financial support from the EPSRC (Grant No. EP/W015641/1). 


\textit{Data availability.} The data and code that support the findings of this work are openly available \cite{code}.


\textit{AI usage statement.} The authors used \emph{Anthropic}'s \emph{Claude (Sonnet 4)} for grammar and spell-checking of the manuscript. All suggestions were reviewed and approved by the authors. Further, it was used to extract experimentally measured lifetimes from a publication, which are plotted in the End Matter (please see the \ac{em} for details).

\bibliography{references}

\begin{figure*}
    \centering
    \includegraphics{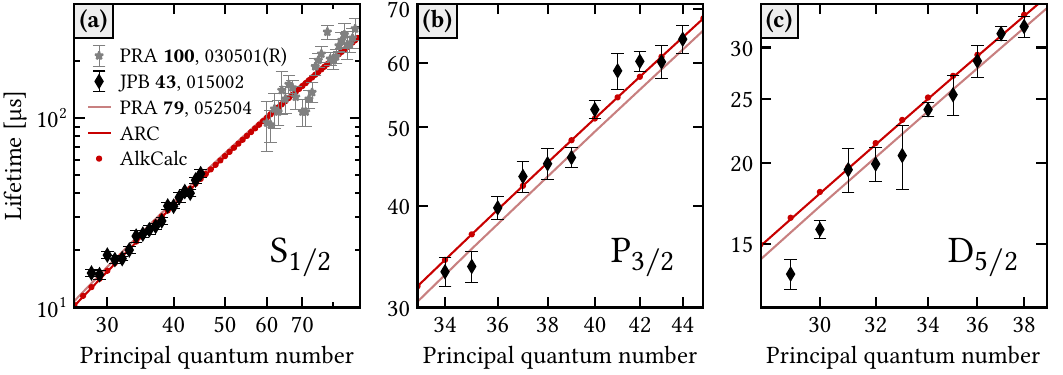}
    \caption{\textbf{Benchmark of room-temperature lifetimes of Rydberg states in $\bm{\Rubidium}$.} (a) S-state lifetimes. Red dots show lifetimes of the Rydberg $n\mathrm{S}_{1\slash{2}}$ states in $\mathrm{^{85}Rb}$ computed with \emph{AlkCalc} \cite{alkcalc}. Black diamonds show experimental results from Ref.~\cite{branden2010} for $\mathrm{^{85}Rb}$. Grey stars show experimental results from Ref.~\cite{archimi2019}. In the latter case, the data shown here is not available in tabulated form. Therefore, we employed \emph{Anthropic}'s \emph{Claude (Sonnet 4)} to extract the data directly from the figure in the portable-document-format (PDF) file. We note that due to this method, the precision of the data and the associated error bars is not quantitatively known. However, the extracted data (grey stars) agrees well with our numerical prediction. We remark that for the extracted data the specific isotope of Rubidium is not specified. The red line shows the lifetimes computed with the software \emph{ARC} \cite{sibalic2017}. Finally, the light red line shows results from Ref.~\cite{beterov2009,beterov2009erratum}. (b) P-state lifetimes. Same plot as in (a) but for the Rydberg states $n\mathrm{P}_{3\slash{2}}$. (c) D-state lifetimes. Same plot as in (a) but for the Rydberg states $n\mathrm{D}_{5\slash{2}}$.
    }
    \label{fig:BenchmarkRoomTemperature}
\end{figure*}

\begin{figure*}
    \centering
    \includegraphics{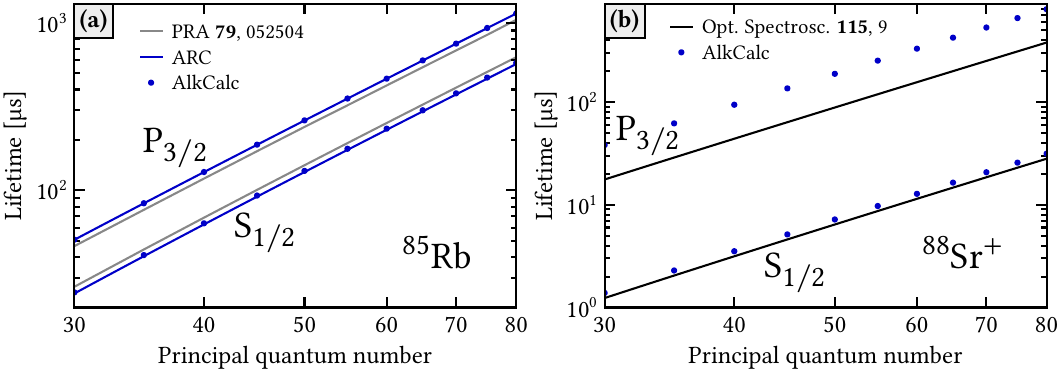}
    \caption{\textbf{Comparison of zero-temperature lifetimes of Rydberg states in $\bm{\Rubidium}$ and $\bm{\Strontium}$ to numerical results from the literature.} (a) Comparison of lifetimes obtained with \emph{AlkCalc} \cite{alkcalc} (blue dots) and \emph{ARC} \cite{sibalic2017} (blue lines) to results from Ref.~\cite{beterov2009,beterov2009erratum} (grey lines). The lifetimes are shown for S and P states in $\Rubidium$. (b) Comparison of lifetimes obtained with \emph{AlkCalc} (blue dots) to results from Ref.~\cite{glukhov2013} (black lines). The lifetimes are shown for S and P states in $\Strontium$.
    }
    \label{fig:BenchmarkZeroTemperature}
\end{figure*}

\section{End Matter}

We benchmark our numerical method against experimental and numerical data from the literature. To our knowledge, reasonably large sets of experimentally measured Rydberg-state lifetimes exist only for the Rydberg states $n\mathrm{S}_{1\slash2}$, $n\mathrm{P}_{3\slash2}$, and $n\mathrm{D}_{5\slash2}$ in $\Rubidium$ at room temperature. Figure~\ref{fig:BenchmarkRoomTemperature} compares these measured lifetimes (black diamonds and grey stars) to the lifetimes predicted by \emph{AlkCalc}~\cite{alkcalc} (red dots), the software used in this work, as well to two other theoretical works from the literature: ARC~\cite{sibalic2017} (red line) and Ref.~\cite{beterov2009,beterov2009erratum} (light red line). All results agree well, except for a slight deviation of the lifetimes obtained in Ref.~\cite{beterov2009,beterov2009erratum} from the predictions by \emph{AlkCalc} and \emph{ARC}. The deviation is visible by comparing the red and light red curves. Further, we benchmark the lifetimes computed with \emph{AlkCalc} at zero temperature against theoretical lifetimes from Ref.~\cite{beterov2009,beterov2009erratum} (grey line) and from ARC~\cite{sibalic2017} (blue line). Figure~\ref{fig:BenchmarkZeroTemperature}(a) shows good agreement between our method, \emph{ARC}, and Ref.~\cite{beterov2009,beterov2009erratum}, with the results from Ref.~\cite{beterov2009,beterov2009erratum} again deviating slightly. Panel~(b) shows the analogous comparison for $\Strontium$. Note that for $\Strontium$ the software \emph{ARC} is unavailable since it is limited to alkali-metal atoms. Instead, we compare our results to the numerical data obtained in Ref.~\cite{glukhov2013}. The S-state lifetimes agree better than the P-state lifetimes, though the anomalous enhancement of the P-state lifetimes remains clearly visible also in the data from Ref.~\cite{glukhov2013} (black lines).

A likely source of the quantitative discrepancy for the zero-temperature P-state lifetimes in $\Strontium$ is the model potential used for computing the eigenenergies and eigenstates: Ref.~\cite{glukhov2013} uses Fuse's model \cite{simons1971}. This model potential, $V$, is given by the potential for Hydrogenic ions with the replacement $l\mapsto\nu_l$:
\begin{equation}
    V(r)
    =
    -
    \frac{Z_\mathrm{eff}e^2}{4\pii\varepsilon_0}\frac{1}{r}
    +
    \frac{\hbar^2\nu_l(\nu_l+1)}{2mr^2}
\,,
\end{equation}
where $\nu_l$ is an $l$-dependent continuous parameter, optimised to reproduce the correct spectroscopic eigenenergies, and $Z_\mathrm{eff}=2$ is the effective nuclear charge. In contrast, the parametric model potential \cite{aymar1996} employed by \emph{AlkCalc} has a more sophisticated structure at small radii $r$ with a larger set of independent parameters, optimised to match spectroscopic data. Therefore, we expect that the radial eigenstates computed with \emph{AlkCalc} capture the complex small-$r$ behaviour of the wave function more accurately than Fuse's model. This matters particularly for zero-temperature lifetimes, since the dominant transition matrix elements are between Rydberg states and low-lying electronic states. The latter typically penetrate the ground-state electron cloud and the nucleus deeply and are therefore highly sensitive to the small-$r$ behaviour of the model potential. In particular for Rydberg P states, where the transition matrix elements show accidental cancellations, precisely reproducing the small-$r$ behaviour of the radial wave function is especially important. Therefore, this explanation aligns well with the fact that the quantitative discrepancy is most pronounced for zero-temperature lifetimes of P states, as suggested by Fig.~\ref{fig:BenchmarkZeroTemperature}(b) (cf. blue dots and black lines).

\clearpage
\setcounter{secnumdepth}{3}
\setcounter{section}{0}
\setcounter{equation}{0}
\setcounter{figure}{0}
\setcounter{table}{0}
\renewcommand{\thesection}{S\arabic{section}}
\renewcommand{\theequation}{S\arabic{equation}}
\renewcommand{\thefigure}{S\arabic{figure}}
\renewcommand{\thetable}{S\arabic{table}}
\onecolumngrid
\begin{center}
        \large\bf
        Supplemental material for:
\\
        Anomalously enhanced lifetimes of low angular momentum Rydberg states in
        singly charged alkaline-earth metal ions
\end{center}
\vspace*{0mm}
\begin{center}
    \begin{minipage}{15cm}
        \begin{center}
            Simon Euchner$^1$, Weibin Li$^2$, and Igor Lesanovsky$^{1,2}$
\vspace*{2mm}
\\
            $^1$\textit{Institut f\"ur Theoretische Physik and Center for Integrated Quantum Science and Technology,
\\
    Universit\"at T\"ubingen, Auf der Morgenstelle 14, 72076 T\"ubingen, Germany}
\\
            $^2$\textit{School of Physics and Astronomy and Centre for the Mathematics
\\
            and Theoretical Physics of Quantum Non-Equilibrium Systems,
\\
            The University of Nottingham, Nottingham, NG7 2RD, United Kingdom}
        \end{center}
    \end{minipage}
\end{center}

\vspace*{0mm}
\begin{center}
    \begin{minipage}{14.5cm}
        \hspace*{1em}  In Sec.~\ref{sec:LifetimesOfRydbergStatesInCalciumStrontiumAndRubidium} we provide tables with numerically computed lifetimes of Rydberg states in $\Calcium$, $\Strontium$, and $\Rubidium$ at zero temperature and room temperature. In Sec.~\ref{sec:FormulasForCalculatingLifetimesOfRydbergStatesInCalciumStrontiumAndRubidium} we provide formulas for calculating Rydberg-state lifetimes in $\Calcium$, $\Strontium$, and $\Rubidium$ 
    \end{minipage}
\end{center}
\vspace*{5mm}

\section{\texorpdfstring{Lifetimes of Rydberg states in ${\bm{\Calcium}}$, $\bm{\Strontium}$, and $\bm{\Rubidium}$}{Lifetimes of Rydberg states in Calcium-40, Strontium-88, and Rubidium-85}}
\label{sec:LifetimesOfRydbergStatesInCalciumStrontiumAndRubidium}

In Tabs.~\ref{tab:LifetimesCalcium}, \ref{tab:LifetimesStrontium}, and \ref{tab:LifetimesRubidium} numerically computed lifetimes are tabulated for the Calcium ion $\Calcium$ and the Strontium ion $\Strontium$, and the Rubidium atom $\Rubidium$, respectively. The lifetimes are computed at zero temperature, yielding the radiative lifetimes, $\tau_\mathrm{rad}$, and at room temperature $T=300\,\mathrm{K}$, $\tau_\mathrm{bbr}$, yielding lifetimes that are reduced by black-body induced transitions. All lifetimes are calculated within the dipole approximation \cite{friedrich2017,alkcalc}. As discussed in the main text, black-body radiation can depopulate Rydberg states by transitions to other Rydberg states that lie \emph{higher} in energy than the level of interest. To account for this numerically, we truncate the electronic basis 30 states above the Rydberg level of interest, i.e., we consider black-body induced excitation of the Rydberg level $n\mathrm{L}_j$ up to and including the Rydberg level $(n+30)\mathrm{L'}_{j'}$.

\begin{table}[ht]
    \centering
    \caption{\textbf{Lifetimes of Rydberg states in $\bm{\Calcium}$.} Lifetimes calculated with \emph{AlkCalc} \cite{alkcalc} for principal quantum numbers $20\leq{n}\leq80$. Values are in microseconds (\textmu{s}) and given to four significant digits. ‘rad’ denotes radiative ($T=0\,\mathrm{K}$) lifetimes, while ‘bbr’ denotes room-temperature ($T=300\,\mathrm{K}$) lifetimes, which are reduced by black-body-radiation-induced transitions.}
    \label{tab:LifetimesCalcium}
    \begin{tabular}{lrrrrrrrrrr}
        \toprule
        {$n$} & {$\mathrm{S_\frac{1}{2}}$(rad)} & {$\mathrm{P_\frac{1}{2}}$(rad)} & {$\mathrm{P_\frac{3}{2}}$(rad)} & {$\mathrm{D_\frac{3}{2}}$(rad)} & {$\mathrm{D_\frac{5}{2}}$(rad)} & {$\mathrm{S_\frac{1}{2}}$(bbr)} & {$\mathrm{P_\frac{1}{2}}$(bbr)} & {$\mathrm{P_\frac{3}{2}}$(bbr)} & {$\mathrm{D_\frac{3}{2}}$(bbr)} & {$\mathrm{D_\frac{5}{2}}$(bbr)} \\
        \midrule
        20 & 0.3882 & 8.472 & 9.013 & 0.5231 & 0.5299 & 0.3703 & 4.507 & 4.637 & 0.4936 & 0.4996 \\
        21 & 0.455 & 9.88 & 10.52 & 0.6085 & 0.6165 & 0.4319 & 5.001 & 5.139 & 0.5711 & 0.578 \\
        22 & 0.5291 & 11.44 & 12.18 & 0.7028 & 0.712 & 0.4997 & 5.52 & 5.666 & 0.6559 & 0.6638 \\
        23 & 0.6109 & 13.14 & 14.01 & 0.8064 & 0.817 & 0.5741 & 6.065 & 6.219 & 0.7485 & 0.7574 \\
        24 & 0.7007 & 15.01 & 16.01 & 0.9197 & 0.9318 & 0.6554 & 6.637 & 6.798 & 0.8489 & 0.8591 \\
        25 & 0.799 & 17.05 & 18.19 & 1.043 & 1.057 & 0.7437 & 7.235 & 7.404 & 0.9577 & 0.969 \\
        26 & 0.9061 & 19.27 & 20.56 & 1.177 & 1.193 & 0.8393 & 7.861 & 8.038 & 1.075 & 1.088 \\
        27 & 1.022 & 21.66 & 23.12 & 1.322 & 1.339 & 0.9425 & 8.515 & 8.699 & 1.201 & 1.215 \\
        28 & 1.148 & 24.25 & 25.89 & 1.478 & 1.498 & 1.054 & 9.197 & 9.389 & 1.336 & 1.351 \\
        29 & 1.284 & 27.04 & 28.87 & 1.647 & 1.668 & 1.173 & 9.908 & 10.11 & 1.48 & 1.497 \\
        30 & 1.43 & 30.03 & 32.07 & 1.827 & 1.851 & 1.3 & 10.65 & 10.85 & 1.634 & 1.653 \\
        31 & 1.587 & 33.23 & 35.5 & 2.02 & 2.047 & 1.436 & 11.42 & 11.63 & 1.797 & 1.818 \\
        32 & 1.755 & 36.66 & 39.16 & 2.227 & 2.256 & 1.58 & 12.21 & 12.44 & 1.971 & 1.994 \\
        33 & 1.935 & 40.31 & 43.07 & 2.447 & 2.479 & 1.734 & 13.04 & 13.27 & 2.155 & 2.179 \\
        34 & 2.126 & 44.19 & 47.23 & 2.681 & 2.716 & 1.896 & 13.9 & 14.14 & 2.349 & 2.376 \\
        35 & 2.329 & 48.32 & 51.65 & 2.929 & 2.968 & 2.069 & 14.79 & 15.03 & 2.554 & 2.583 \\
        36 & 2.546 & 52.7 & 56.33 & 3.193 & 3.235 & 2.25 & 15.7 & 15.96 & 2.77 & 2.801 \\
        37 & 2.775 & 57.33 & 61.29 & 3.471 & 3.517 & 2.442 & 16.65 & 16.91 & 2.997 & 3.03 \\
        38 & 3.017 & 62.23 & 66.53 & 3.766 & 3.816 & 2.643 & 17.63 & 17.9 & 3.235 & 3.271 \\
        39 & 3.274 & 67.4 & 72.06 & 4.077 & 4.131 & 2.855 & 18.64 & 18.92 & 3.485 & 3.524 \\
        40 & 3.544 & 72.85 & 77.9 & 4.404 & 4.462 & 3.077 & 19.67 & 19.96 & 3.747 & 3.788 \\
        41 & 3.829 & 78.59 & 84.03 & 4.748 & 4.811 & 3.309 & 20.74 & 21.04 & 4.02 & 4.064 \\
        42 & 4.129 & 84.62 & 90.49 & 5.11 & 5.178 & 3.553 & 21.84 & 22.15 & 4.306 & 4.353 \\
        43 & 4.444 & 90.95 & 97.26 & 5.49 & 5.563 & 3.807 & 22.97 & 23.29 & 4.604 & 4.654 \\
        44 & 4.775 & 97.58 & 104.4 & 5.888 & 5.966 & 4.072 & 24.13 & 24.45 & 4.915 & 4.968 \\
        45 & 5.122 & 104.5 & 111.8 & 6.305 & 6.389 & 4.349 & 25.32 & 25.65 & 5.238 & 5.295 \\
        46 & 5.485 & 111.8 & 119.6 & 6.741 & 6.831 & 4.637 & 26.54 & 26.88 & 5.575 & 5.635 \\
        47 & 5.865 & 119.4 & 127.7 & 7.197 & 7.292 & 4.937 & 27.8 & 28.14 & 5.924 & 5.988 \\
        48 & 6.263 & 127.4 & 136.2 & 7.673 & 7.775 & 5.248 & 29.08 & 29.43 & 6.287 & 6.354 \\
        49 & 6.677 & 135.7 & 145.1 & 8.169 & 8.278 & 5.572 & 30.39 & 30.75 & 6.664 & 6.734 \\
        50 & 7.11 & 144.3 & 154.4 & 8.687 & 8.802 & 5.908 & 31.74 & 32.11 & 7.053 & 7.128 \\
        51 & 7.561 & 153.3 & 164 & 9.225 & 9.348 & 6.256 & 33.11 & 33.49 & 7.457 & 7.535 \\
        52 & 8.031 & 162.7 & 174 & 9.786 & 9.916 & 6.616 & 34.52 & 34.9 & 7.875 & 7.957 \\
        53 & 8.52 & 172.4 & 184.5 & 10.37 & 10.51 & 6.989 & 35.95 & 36.35 & 8.307 & 8.393 \\
        54 & 9.028 & 182.6 & 195.3 & 10.97 & 11.12 & 7.375 & 37.42 & 37.82 & 8.753 & 8.844 \\
        55 & 9.556 & 193.1 & 206.6 & 11.6 & 11.76 & 7.774 & 38.92 & 39.33 & 9.214 & 9.309 \\
        56 & 10.1 & 204 & 218.2 & 12.26 & 12.42 & 8.186 & 40.44 & 40.86 & 9.689 & 9.788 \\
        57 & 10.67 & 215.3 & 230.4 & 12.93 & 13.1 & 8.611 & 42 & 42.43 & 10.18 & 10.28 \\
        58 & 11.26 & 227 & 242.9 & 13.63 & 13.81 & 9.049 & 43.59 & 44.03 & 10.68 & 10.79 \\
        59 & 11.88 & 239.2 & 255.9 & 14.36 & 14.55 & 9.501 & 45.21 & 45.66 & 11.2 & 11.32 \\
        60 & 12.51 & 251.8 & 269.4 & 15.11 & 15.31 & 9.966 & 46.86 & 47.32 & 11.74 & 11.86 \\
        61 & 13.16 & 264.8 & 283.3 & 15.89 & 16.1 & 10.45 & 48.55 & 49.01 & 12.29 & 12.41 \\
        62 & 13.84 & 278.2 & 297.7 & 16.69 & 16.91 & 10.94 & 50.26 & 50.73 & 12.86 & 12.98 \\
        63 & 14.54 & 292.1 & 312.6 & 17.52 & 17.75 & 11.45 & 52 & 52.48 & 13.44 & 13.57 \\
        64 & 15.27 & 306.5 & 327.9 & 18.37 & 18.62 & 11.97 & 53.78 & 54.26 & 14.04 & 14.17 \\
        65 & 16.01 & 321.3 & 343.8 & 19.26 & 19.51 & 12.5 & 55.58 & 56.07 & 14.65 & 14.79 \\
        66 & 16.79 & 336.6 & 360.2 & 20.17 & 20.44 & 13.05 & 57.42 & 57.92 & 15.28 & 15.43 \\
        67 & 17.58 & 352.4 & 377 & 21.11 & 21.39 & 13.62 & 59.28 & 59.79 & 15.92 & 16.08 \\
        68 & 18.4 & 368.6 & 394.4 & 22.08 & 22.37 & 14.2 & 61.18 & 61.7 & 16.59 & 16.75 \\
        69 & 19.25 & 385.4 & 412.4 & 23.08 & 23.38 & 14.79 & 63.11 & 63.63 & 17.27 & 17.43 \\
        70 & 20.12 & 402.6 & 430.8 & 24.1 & 24.42 & 15.4 & 65.06 & 65.6 & 17.96 & 18.13 \\
        71 & 21.02 & 420.4 & 449.8 & 25.16 & 25.5 & 16.02 & 67.05 & 67.59 & 18.67 & 18.85 \\
        72 & 21.94 & 438.6 & 469.4 & 26.25 & 26.6 & 16.66 & 69.07 & 69.62 & 19.4 & 19.59 \\
        73 & 22.89 & 457.4 & 489.5 & 27.37 & 27.73 & 17.31 & 71.12 & 71.68 & 20.15 & 20.34 \\
        74 & 23.87 & 476.7 & 510.2 & 28.52 & 28.9 & 17.98 & 73.21 & 73.77 & 20.91 & 21.11 \\
        75 & 24.87 & 496.6 & 531.4 & 29.7 & 30.09 & 18.67 & 75.32 & 75.89 & 21.69 & 21.89 \\
        76 & 25.91 & 517 & 553.3 & 30.91 & 31.33 & 19.37 & 77.46 & 78.04 & 22.49 & 22.7 \\
        77 & 26.97 & 538 & 575.7 & 32.16 & 32.59 & 20.08 & 79.63 & 80.22 & 23.3 & 23.52 \\
        78 & 28.06 & 559.5 & 598.7 & 33.44 & 33.89 & 20.82 & 81.84 & 82.43 & 24.13 & 24.36 \\
        79 & 29.18 & 581.6 & 622.4 & 34.76 & 35.22 & 21.56 & 84.07 & 84.67 & 24.98 & 25.21 \\
        80 & 30.32 & 604.2 & 646.6 & 36.1 & 36.58 & 22.33 & 86.34 & 86.95 & 25.85 & 26.09 \\
        \bottomrule
    \end{tabular}
\end{table}

\begin{table}[ht]
    \centering
    \caption{\textbf{Lifetimes of Rydberg states in $\bm{\Strontium}$.} Lifetimes calculated with \emph{AlkCalc} \cite{alkcalc} for principal quantum numbers $20\leq{n}\leq80$. Values are in microseconds (\textmu{s}) and given to four significant digits. ‘rad’ denotes radiative ($T=0\,\mathrm{K}$) lifetimes, while ‘bbr’ denotes room-temperature ($T=300\,\mathrm{K}$) lifetimes, which are reduced by black-body-radiation-induced transitions.
    }
    \label{tab:LifetimesStrontium}
    \begin{tabular}{lrrrrrrrrrr}
        \toprule
        {$n$} & {$\mathrm{S_\frac{1}{2}}$(rad)} & {$\mathrm{P_\frac{1}{2}}$(rad)} & {$\mathrm{P_\frac{3}{2}}$(rad)} & {$\mathrm{D_\frac{3}{2}}$(rad)} & {$\mathrm{D_\frac{5}{2}}$(rad)} & {$\mathrm{S_\frac{1}{2}}$(bbr)} & {$\mathrm{P_\frac{1}{2}}$(bbr)} & {$\mathrm{P_\frac{3}{2}}$(bbr)} & {$\mathrm{D_\frac{3}{2}}$(bbr)} & {$\mathrm{D_\frac{5}{2}}$(bbr)} \\
        \midrule
        20 & 0.3563 & 8.984 & 10.31 & 0.3677 & 0.3798 & 0.3397 & 4.745 & 5.009 & 0.3542 & 0.3654 \\
        21 & 0.4209 & 10.5 & 12.1 & 0.4299 & 0.4441 & 0.3992 & 5.25 & 5.531 & 0.4124 & 0.4254 \\
        22 & 0.493 & 12.19 & 14.09 & 0.4989 & 0.5153 & 0.4651 & 5.778 & 6.077 & 0.4766 & 0.4915 \\
        23 & 0.573 & 14.04 & 16.27 & 0.5748 & 0.5938 & 0.5377 & 6.332 & 6.647 & 0.5469 & 0.5639 \\
        24 & 0.6612 & 16.06 & 18.67 & 0.6581 & 0.6798 & 0.6172 & 6.912 & 7.243 & 0.6235 & 0.6428 \\
        25 & 0.758 & 18.28 & 21.29 & 0.7491 & 0.7738 & 0.7039 & 7.518 & 7.865 & 0.7068 & 0.7286 \\
        26 & 0.8639 & 20.68 & 24.14 & 0.8482 & 0.8761 & 0.798 & 8.152 & 8.514 & 0.7969 & 0.8214 \\
        27 & 0.9793 & 23.29 & 27.23 & 0.9556 & 0.9871 & 0.9 & 8.814 & 9.191 & 0.8941 & 0.9214 \\
        28 & 1.105 & 26.11 & 30.57 & 1.072 & 1.107 & 1.01 & 9.504 & 9.897 & 0.9986 & 1.029 \\
        29 & 1.24 & 29.15 & 34.17 & 1.197 & 1.236 & 1.128 & 10.22 & 10.63 & 1.111 & 1.144 \\
        30 & 1.386 & 32.42 & 38.04 & 1.332 & 1.376 & 1.255 & 10.97 & 11.39 & 1.23 & 1.268 \\
        31 & 1.544 & 35.92 & 42.18 & 1.476 & 1.525 & 1.39 & 11.75 & 12.19 & 1.358 & 1.399 \\
        32 & 1.712 & 39.67 & 46.62 & 1.63 & 1.684 & 1.534 & 12.56 & 13.01 & 1.494 & 1.539 \\
        33 & 1.893 & 43.67 & 51.36 & 1.795 & 1.855 & 1.688 & 13.39 & 13.86 & 1.639 & 1.687 \\
        34 & 2.086 & 47.93 & 56.41 & 1.971 & 2.036 & 1.851 & 14.26 & 14.75 & 1.791 & 1.845 \\
        35 & 2.292 & 52.46 & 61.78 & 2.158 & 2.229 & 2.023 & 15.16 & 15.66 & 1.953 & 2.011 \\
        36 & 2.511 & 57.27 & 67.48 & 2.356 & 2.434 & 2.206 & 16.09 & 16.6 & 2.124 & 2.187 \\
        37 & 2.743 & 62.36 & 73.52 & 2.566 & 2.651 & 2.398 & 17.05 & 17.57 & 2.304 & 2.372 \\
        38 & 2.99 & 67.75 & 79.91 & 2.788 & 2.88 & 2.601 & 18.04 & 18.58 & 2.493 & 2.566 \\
        39 & 3.25 & 73.44 & 86.66 & 3.023 & 3.123 & 2.814 & 19.05 & 19.61 & 2.692 & 2.771 \\
        40 & 3.526 & 79.44 & 93.77 & 3.27 & 3.378 & 3.038 & 20.1 & 20.68 & 2.901 & 2.985 \\
        41 & 3.816 & 85.77 & 101.3 & 3.531 & 3.647 & 3.273 & 21.19 & 21.77 & 3.12 & 3.21 \\
        42 & 4.123 & 92.42 & 109.2 & 3.805 & 3.931 & 3.519 & 22.3 & 22.9 & 3.349 & 3.445 \\
        43 & 4.445 & 99.4 & 117.4 & 4.093 & 4.228 & 3.776 & 23.44 & 24.06 & 3.588 & 3.69 \\
        44 & 4.783 & 106.7 & 126.1 & 4.395 & 4.54 & 4.045 & 24.61 & 25.25 & 3.838 & 3.947 \\
        45 & 5.139 & 114.4 & 135.2 & 4.712 & 4.867 & 4.325 & 25.82 & 26.47 & 4.098 & 4.214 \\
        46 & 5.511 & 122.5 & 144.8 & 5.044 & 5.21 & 4.617 & 27.05 & 27.72 & 4.369 & 4.492 \\
        47 & 5.901 & 130.9 & 154.8 & 5.391 & 5.568 & 4.921 & 28.32 & 29 & 4.651 & 4.782 \\
        48 & 6.309 & 139.7 & 165.2 & 5.753 & 5.942 & 5.238 & 29.61 & 30.31 & 4.945 & 5.083 \\
        49 & 6.736 & 148.9 & 176.1 & 6.131 & 6.333 & 5.566 & 30.94 & 31.65 & 5.249 & 5.395 \\
        50 & 7.181 & 158.4 & 187.5 & 6.526 & 6.741 & 5.907 & 32.3 & 33.02 & 5.566 & 5.72 \\
        51 & 7.646 & 168.4 & 199.3 & 6.937 & 7.165 & 6.261 & 33.69 & 34.43 & 5.894 & 6.056 \\
        52 & 8.13 & 178.8 & 211.6 & 7.365 & 7.607 & 6.628 & 35.11 & 35.86 & 6.233 & 6.404 \\
        53 & 8.635 & 189.6 & 224.5 & 7.81 & 8.067 & 7.007 & 36.56 & 37.32 & 6.585 & 6.765 \\
        54 & 9.159 & 200.9 & 237.8 & 8.273 & 8.545 & 7.4 & 38.04 & 38.82 & 6.949 & 7.138 \\
        55 & 9.705 & 212.6 & 251.6 & 8.754 & 9.042 & 7.806 & 39.55 & 40.35 & 7.325 & 7.523 \\
        56 & 10.27 & 224.7 & 266 & 9.253 & 9.557 & 8.226 & 41.09 & 41.9 & 7.713 & 7.921 \\
        57 & 10.86 & 237.3 & 280.9 & 9.77 & 10.09 & 8.659 & 42.67 & 43.49 & 8.114 & 8.332 \\
        58 & 11.47 & 250.3 & 296.4 & 10.31 & 10.65 & 9.106 & 44.27 & 45.11 & 8.528 & 8.756 \\
        59 & 12.1 & 263.8 & 312.5 & 10.86 & 11.22 & 9.566 & 45.91 & 46.76 & 8.954 & 9.193 \\
        60 & 12.76 & 277.8 & 329.1 & 11.44 & 11.82 & 10.04 & 47.57 & 48.44 & 9.394 & 9.643 \\
        61 & 13.44 & 292.3 & 346.2 & 12.03 & 12.43 & 10.53 & 49.27 & 50.15 & 9.847 & 10.11 \\
        62 & 14.14 & 307.3 & 364 & 12.65 & 13.07 & 11.03 & 51 & 51.89 & 10.31 & 10.58 \\
        63 & 14.87 & 322.8 & 382.4 & 13.29 & 13.72 & 11.55 & 52.76 & 53.67 & 10.79 & 11.07 \\
        64 & 15.62 & 338.8 & 401.3 & 13.94 & 14.4 & 12.08 & 54.55 & 55.47 & 11.28 & 11.58 \\
        65 & 16.4 & 355.3 & 420.9 & 14.62 & 15.1 & 12.63 & 56.37 & 57.3 & 11.79 & 12.1 \\
        66 & 17.2 & 372.3 & 441.2 & 15.32 & 15.83 & 13.19 & 58.22 & 59.17 & 12.31 & 12.63 \\
        67 & 18.03 & 389.9 & 462 & 16.05 & 16.58 & 13.77 & 60.1 & 61.06 & 12.84 & 13.18 \\
        68 & 18.88 & 408 & 483.5 & 16.79 & 17.34 & 14.36 & 62.01 & 62.99 & 13.39 & 13.74 \\
        69 & 19.76 & 426.7 & 505.7 & 17.56 & 18.14 & 14.97 & 63.95 & 64.94 & 13.96 & 14.31 \\
        70 & 20.67 & 446 & 528.5 & 18.35 & 18.95 & 15.59 & 65.93 & 66.93 & 14.53 & 14.9 \\
        71 & 21.6 & 465.8 & 552.1 & 19.17 & 19.8 & 16.23 & 67.93 & 68.95 & 15.12 & 15.51 \\
        72 & 22.56 & 486.2 & 576.3 & 20 & 20.66 & 16.88 & 69.96 & 70.99 & 15.73 & 16.12 \\
        73 & 23.55 & 507.2 & 601.2 & 20.87 & 21.55 & 17.55 & 72.03 & 73.07 & 16.35 & 16.76 \\
        74 & 24.57 & 528.8 & 626.8 & 21.75 & 22.47 & 18.23 & 74.13 & 75.18 & 16.98 & 17.41 \\
        75 & 25.62 & 551 & 653.1 & 22.66 & 23.41 & 18.93 & 76.25 & 77.32 & 17.63 & 18.07 \\
        76 & 26.7 & 573.8 & 680.2 & 23.6 & 24.38 & 19.65 & 78.41 & 79.49 & 18.3 & 18.75 \\
        77 & 27.81 & 597.2 & 708 & 24.56 & 25.37 & 20.38 & 80.6 & 81.69 & 18.98 & 19.44 \\
        78 & 28.94 & 621.3 & 736.5 & 25.55 & 26.39 & 21.13 & 82.82 & 83.92 & 19.67 & 20.15 \\
        79 & 30.11 & 646 & 765.8 & 26.56 & 27.44 & 21.9 & 85.07 & 86.19 & 20.38 & 20.88 \\
        80 & 31.31 & 671.4 & 795.9 & 27.6 & 28.51 & 22.68 & 87.35 & 88.48 & 21.1 & 21.62 \\
        \bottomrule
    \end{tabular}
\end{table}

\begin{table}[ht]
    \centering
    \caption{\textbf{Lifetimes of Rydberg states in $\bm{\Rubidium}$.} Lifetimes calculated with \emph{AlkCalc} \cite{alkcalc} for principal quantum numbers $20\leq{n}\leq80$. Values are in microseconds (\textmu{s}) and given to four significant digits. ‘rad’ denotes radiative ($T=0\,\mathrm{K}$) lifetimes, while ‘bbr’ denotes room-temperature ($T=300\,\mathrm{K}$) lifetimes, which are reduced by black-body-radiation-induced transitions.}
    \label{tab:LifetimesRubidium}
    \begin{tabular}{lrrrrrrrrrr}
        \toprule
        {$n$} & {$\mathrm{S_\frac{1}{2}}$(rad)} & {$\mathrm{P_\frac{1}{2}}$(rad)} & {$\mathrm{P_\frac{3}{2}}$(rad)} & {$\mathrm{D_\frac{3}{2}}$(rad)} & {$\mathrm{D_\frac{5}{2}}$(rad)} & {$\mathrm{S_\frac{1}{2}}$(bbr)} & {$\mathrm{P_\frac{1}{2}}$(bbr)} & {$\mathrm{P_\frac{3}{2}}$(bbr)} & {$\mathrm{D_\frac{3}{2}}$(bbr)} & {$\mathrm{D_\frac{5}{2}}$(bbr)} \\
        \midrule
        20 & 6.104 & 14.12 & 13.16 & 7.49 & 7.389 & 4.633 & 8.827 & 8.467 & 5.944 & 5.891 \\
        21 & 7.252 & 16.66 & 15.53 & 8.716 & 8.599 & 5.397 & 10.1 & 9.696 & 6.809 & 6.75 \\
        22 & 8.537 & 19.49 & 18.17 & 10.07 & 9.936 & 6.231 & 11.46 & 11.02 & 7.747 & 7.68 \\
        23 & 9.964 & 22.63 & 21.1 & 11.56 & 11.41 & 7.138 & 12.91 & 12.43 & 8.758 & 8.685 \\
        24 & 11.54 & 26.09 & 24.32 & 13.2 & 13.02 & 8.119 & 14.47 & 13.95 & 9.846 & 9.765 \\
        25 & 13.28 & 29.88 & 27.87 & 14.98 & 14.78 & 9.175 & 16.11 & 15.55 & 11.01 & 10.92 \\
        26 & 15.19 & 34.03 & 31.74 & 16.92 & 16.69 & 10.31 & 17.86 & 17.25 & 12.25 & 12.16 \\
        27 & 17.27 & 38.55 & 35.95 & 19.02 & 18.76 & 11.52 & 19.7 & 19.05 & 13.58 & 13.47 \\
        28 & 19.53 & 43.45 & 40.52 & 21.29 & 21 & 12.81 & 21.64 & 20.94 & 14.98 & 14.86 \\
        29 & 21.98 & 48.76 & 45.47 & 23.74 & 23.41 & 14.18 & 23.68 & 22.93 & 16.46 & 16.34 \\
        30 & 24.62 & 54.48 & 50.8 & 26.36 & 26 & 15.63 & 25.81 & 25.02 & 18.03 & 17.9 \\
        31 & 27.48 & 60.63 & 56.54 & 29.18 & 28.78 & 17.16 & 28.04 & 27.2 & 19.68 & 19.54 \\
        32 & 30.54 & 67.23 & 62.7 & 32.19 & 31.75 & 18.78 & 30.37 & 29.48 & 21.42 & 21.27 \\
        33 & 33.83 & 74.29 & 69.28 & 35.4 & 34.92 & 20.48 & 32.8 & 31.86 & 23.24 & 23.08 \\
        34 & 37.34 & 81.84 & 76.32 & 38.82 & 38.29 & 22.27 & 35.33 & 34.34 & 25.15 & 24.98 \\
        35 & 41.09 & 89.88 & 83.82 & 42.46 & 41.88 & 24.14 & 37.96 & 36.92 & 27.15 & 26.96 \\
        36 & 45.08 & 98.43 & 91.79 & 46.32 & 45.68 & 26.09 & 40.69 & 39.59 & 29.23 & 29.04 \\
        37 & 49.32 & 107.5 & 100.3 & 50.4 & 49.71 & 28.14 & 43.51 & 42.37 & 31.41 & 31.2 \\
        38 & 53.82 & 117.1 & 109.2 & 54.72 & 53.97 & 30.27 & 46.44 & 45.24 & 33.67 & 33.45 \\
        39 & 58.58 & 127.3 & 118.7 & 59.28 & 58.46 & 32.49 & 49.46 & 48.21 & 36.02 & 35.79 \\
        40 & 63.62 & 138.1 & 128.7 & 64.09 & 63.21 & 34.8 & 52.59 & 51.28 & 38.46 & 38.22 \\
        41 & 68.94 & 149.4 & 139.3 & 69.15 & 68.2 & 37.2 & 55.81 & 54.45 & 41 & 40.74 \\
        42 & 74.55 & 161.3 & 150.5 & 74.48 & 73.45 & 39.69 & 59.13 & 57.72 & 43.63 & 43.36 \\
        43 & 80.45 & 173.9 & 162.2 & 80.07 & 78.96 & 42.27 & 62.56 & 61.09 & 46.34 & 46.06 \\
        44 & 86.66 & 187.1 & 174.5 & 85.93 & 84.74 & 44.94 & 66.08 & 64.55 & 49.16 & 48.86 \\
        45 & 93.18 & 201 & 187.4 & 92.08 & 90.81 & 47.7 & 69.7 & 68.12 & 52.06 & 51.76 \\
        46 & 100 & 215.5 & 201 & 98.51 & 97.15 & 50.55 & 73.43 & 71.79 & 55.06 & 54.74 \\
        47 & 107.2 & 230.7 & 215.2 & 105.2 & 103.8 & 53.5 & 77.25 & 75.55 & 58.15 & 57.82 \\
        48 & 114.7 & 246.7 & 230 & 112.3 & 110.7 & 56.53 & 81.17 & 79.42 & 61.34 & 60.99 \\
        49 & 122.5 & 263.3 & 245.5 & 119.6 & 117.9 & 59.66 & 85.2 & 83.39 & 64.62 & 64.26 \\
        50 & 130.7 & 280.7 & 261.7 & 127.2 & 125.5 & 62.89 & 89.32 & 87.45 & 68 & 67.63 \\
        51 & 139.3 & 298.8 & 278.6 & 135.2 & 133.3 & 66.2 & 93.54 & 91.62 & 71.47 & 71.09 \\
        52 & 148.2 & 317.6 & 296.2 & 143.5 & 141.5 & 69.61 & 97.87 & 95.88 & 75.04 & 74.64 \\
        53 & 157.5 & 337.3 & 314.5 & 152.1 & 150 & 73.12 & 102.3 & 100.2 & 78.7 & 78.29 \\
        54 & 167.1 & 357.8 & 333.6 & 161.1 & 158.9 & 76.72 & 106.8 & 104.7 & 82.46 & 82.04 \\
        55 & 177.2 & 379 & 353.4 & 170.4 & 168 & 80.41 & 111.4 & 109.3 & 86.32 & 85.88 \\
        56 & 187.6 & 401.1 & 374 & 180.1 & 177.6 & 84.2 & 116.2 & 113.9 & 90.28 & 89.83 \\
        57 & 198.5 & 424.1 & 395.4 & 190.1 & 187.5 & 88.09 & 121 & 118.7 & 94.33 & 93.86 \\
        58 & 209.7 & 447.9 & 417.6 & 200.5 & 197.7 & 92.07 & 125.9 & 123.6 & 98.48 & 98 \\
        59 & 221.4 & 472.5 & 440.6 & 211.3 & 208.3 & 96.14 & 130.9 & 128.5 & 102.7 & 102.2 \\
        60 & 233.5 & 498.1 & 464.4 & 222.4 & 219.3 & 100.3 & 136.1 & 133.6 & 107.1 & 106.6 \\
        61 & 246.1 & 524.5 & 489.1 & 233.9 & 230.7 & 104.6 & 141.3 & 138.8 & 111.5 & 111 \\
        62 & 259.1 & 551.9 & 514.6 & 245.9 & 242.4 & 108.9 & 146.6 & 144 & 116.1 & 115.5 \\
        63 & 272.5 & 580.3 & 541 & 258.2 & 254.6 & 113.4 & 152.1 & 149.4 & 120.7 & 120.2 \\
        64 & 286.4 & 609.5 & 568.3 & 270.9 & 267.1 & 118 & 157.6 & 154.9 & 125.5 & 124.9 \\
        65 & 300.7 & 639.8 & 596.5 & 284.1 & 280.1 & 122.6 & 163.2 & 160.4 & 130.3 & 129.7 \\
        66 & 315.5 & 671 & 625.7 & 297.6 & 293.5 & 127.4 & 169 & 166.1 & 135.2 & 134.6 \\
        67 & 330.8 & 703.3 & 655.7 & 311.6 & 307.2 & 132.2 & 174.8 & 171.9 & 140.3 & 139.7 \\
        68 & 346.6 & 736.5 & 686.7 & 326 & 321.5 & 137.2 & 180.7 & 177.8 & 145.4 & 144.8 \\
        69 & 362.9 & 770.8 & 718.7 & 340.9 & 336.1 & 142.2 & 186.8 & 183.7 & 150.7 & 150 \\
        70 & 379.7 & 806.1 & 751.6 & 356.2 & 351.2 & 147.4 & 192.9 & 189.8 & 156 & 155.3 \\
        71 & 397 & 842.5 & 785.6 & 371.9 & 366.7 & 152.6 & 199.1 & 196 & 161.4 & 160.8 \\
        72 & 414.8 & 880 & 820.5 & 388.1 & 382.7 & 157.9 & 205.5 & 202.2 & 167 & 166.3 \\
        73 & 433.1 & 918.6 & 856.5 & 404.8 & 399.1 & 163.4 & 211.9 & 208.6 & 172.6 & 171.9 \\
        74 & 452 & 958.3 & 893.5 & 422 & 416 & 168.9 & 218.4 & 215.1 & 178.4 & 177.6 \\
        75 & 471.4 & 999.1 & 931.5 & 439.6 & 433.4 & 174.5 & 225.1 & 221.6 & 184.2 & 183.5 \\
        76 & 491.4 & 1041 & 970.6 & 457.7 & 451.3 & 180.3 & 231.8 & 228.3 & 190.1 & 189.4 \\
        77 & 511.9 & 1084 & 1011 & 476.3 & 469.6 & 186.1 & 238.6 & 235.1 & 196.2 & 195.4 \\
        78 & 533 & 1128 & 1052 & 495.4 & 488.4 & 192 & 245.6 & 242 & 202.3 & 201.6 \\
        79 & 554.6 & 1174 & 1095 & 515 & 507.8 & 198 & 252.6 & 248.9 & 208.6 & 207.8 \\
        80 & 576.8 & 1221 & 1138 & 535.1 & 527.6 & 204.2 & 259.8 & 256 & 214.9 & 214.1 \\
        \bottomrule
    \end{tabular}
\end{table}

\section{Formulas for calculating lifetimes of Rydberg states}
\label{sec:FormulasForCalculatingLifetimesOfRydbergStatesInCalciumStrontiumAndRubidium}

In this section we give simple formulas for calculating the Rydberg-state lifetimes of $\Calcium$, $\Strontium$, and $\Rubidium$ given in Tabs.~\ref{tab:LifetimesCalcium}, \ref{tab:LifetimesStrontium}, and \ref{tab:LifetimesRubidium}. Following Ref.~\cite{beterov2009}, the equation for the radiative (zero-temperature) lifetime is
\begin{equation}\label{eq:ModelRadiativeLifetime}
    \tau^{\mathrm{model}}_\mathrm{rad}[\mathrm{ns}]
    =
    \tau_s[\mathrm{ns}] \times n^\delta_\mathrm{eff}
\,.
\end{equation}
Here, $n_\mathrm{eff}=n-\delta_{l,j}$ is the effective principal quantum number with the quantum defect $\delta_{l,j}$ \cite{friedrich2017}. The quantities $\tau_s$ and $\delta$ are fitting parameters. These parameters are obtained by fitting formula \eqref{eq:ModelRadiativeLifetime} to the tabulated lifetimes. For the lifetimes reduced by black-body radiation at temperature $T=300\,\mathrm{K}$, Ref.~\cite{beterov2009} states the formula
\begin{equation}\label{eq:ModelBBRLifetime}
    \tau^{\mathrm{model}}_\mathrm{bbr}[\mathrm{ns}]
    =
    \bigg(
        \frac{1}{\tau_\mathrm{rad}[\mathrm{ns}]}
        +
        \frac{A[\mathrm{ns}^{-1}]}{n^D_\mathrm{eff}}
        \frac{21.4}{\exp(315780\times{B} \slash (n^C_\mathrm{eff} \times T[\mathrm{K}]))-1}
    \bigg)^{-1}
\,.
\end{equation}
Here, $A$, $B$, $C$, and $D$ are fitting parameters. To compute these fitting parameters we apply the following procedure: we first compute $\tau_s$ and $\delta$ by fitting Eq.~\eqref{eq:ModelRadiativeLifetime} to the radiative lifetimes. Subsequently, we plug the resulting $\tau_s$ and $\delta$ into Eq.~\eqref{eq:ModelBBRLifetime} to reduce $\tau_\mathrm{bbr}=\tau_\mathrm{bbr}(A,B,C,D)$ to a sole function of the fitting parameters $A$, $B$, $C$, and $D$. By fitting this model to the lifetimes reduced by black-body radiation, we obtain the optimal values for $A$, $B$, $C$, and $D$. For the fit itself, we choose a least-squares cost function and minimize it. The results for the fitting parameters and the employed values for the quantum defects, $\delta_{l,j}$, are given in Tab.~\ref{tab:FittingParameters}.

To quantify the quality of the fits, and thereby the formulas \eqref{eq:ModelRadiativeLifetime} and \eqref{eq:ModelBBRLifetime}, we consider the maximal relative errors
\begin{equation}\label{eq:RelativeErrorRadiativeLifetimes}
    \Delta_\mathrm{rad}
    =
    \underset{20\leq{n}\leq80}{\mathrm{max}}
    \frac{\abs*{\tau^\mathrm{model}_\mathrm{rad}(n)-\tau_\mathrm{rad}(n)}}{\tau_\mathrm{rad}(n)}
\end{equation}
and
\begin{equation}\label{eq:RelativeErrorBBRLifetimes} 
    \Delta_\mathrm{bbr}
    =
    \underset{20\leq{n}\leq80}{\mathrm{max}}
    \frac{\abs*{\tau^\mathrm{model}_\mathrm{bbr}(n)-\tau_\mathrm{bbr}(n)}}{\tau_\mathrm{bbr}(n)}
\,,
\end{equation}
for the parameters $\tau_s$, $\delta$, $A$, $B$, $C$, and $D$ returned by the fit and listed in Tab.~\ref{tab:FittingParameters}. These formulas give the `worst-case' relative error when the formulas \eqref{eq:ModelRadiativeLifetime} and \eqref{eq:ModelBBRLifetime} are compared to the numerically obtained lifetimes $\tau_\mathrm{rad}$ and $\tau_\mathrm{bbr}$ given in Tabs.~\ref{tab:LifetimesCalcium}, \ref{tab:LifetimesStrontium}, and \ref{tab:LifetimesRubidium}.

\begin{table}[ht]
    \centering
    \caption{\textbf{Results for fitting parameters.} Fitting parameters for calculating Rydberg-state lifetimes with formulas \eqref{eq:ModelRadiativeLifetime} and \eqref{eq:ModelBBRLifetime}. $\Delta_\mathrm{rad}$ and $\Delta_\mathrm{bbr}$ are the maximal relative errors for the radiative lifetimes ($T=0\,\mathrm{K}$) and the lifetimes reduced by black-body radiation at room temperature ($T=300\,\mathrm{K}$), respectively. Please see Eqs.~\eqref{eq:RelativeErrorRadiativeLifetimes} and \eqref{eq:RelativeErrorBBRLifetimes} for their definition. The quantum defects, $\delta_{l,j}$, are taken from Refs.~\cite{li2003,pokorny2019,mokhberi2019,andrijauskas2021} and are \emph{not} determined by the fit. For $\Strontium$ and $\Calcium$ we use the same quantum defects for both $j=\abs*{l-1\slash{2}}$ and $j=l+1\slash{2}$. This is justified by the quality of the fits (see results for $\Delta_\mathrm{rad}$ and $\Delta_\mathrm{bbr}$).}
    \label{tab:FittingParameters}
    \setlength{\tabcolsep}{6pt}
    \begin{tabular}{llrrrrrrrcc}
        \toprule
        & & {$\delta_{l,j}$} & {$\tau_s[\mathrm{ns}]$} & {$\delta$} & {$A[\mathrm{ns}^{-1}]$} & {$B$} & {$C$} & {$D$} & {$\Delta_\mathrm{rad}$[\textperthousand]} & {$\Delta_\mathrm{bbr}$[\textperthousand]} \\
        \midrule
        {$\Calcium$}   & {$\mathrm{S_\frac{1}{2}}$} & 1.802995 & 0.0643439 & 2.99664 & 1.71505 & 0.635144 & 2.17461 & 4.10589 & 11 & 2 \\
                       & {$\mathrm{P_\frac{1}{2}}$} & 1.4369 & 1.38617 & 2.97545 & 1.07581 & 0.36716 & 2.01395 & 4.00075 & 25 & 3 \\
                       & {$\mathrm{P_\frac{3}{2}}$} & 1.4369 & 1.47706 & 2.97644 & 1.00293 & 0.338701 & 1.99146 & 3.97992 & 21 & 3 \\
                       & {$\mathrm{D_\frac{3}{2}}$} & 0.626888 & 0.071956 & 3.00077 & 0.223293 & 0.0627589 & 1.43837 & 3.43424 & 2 & {$(<1)$} \\
                       & {$\mathrm{D_\frac{5}{2}}$} & 0.626888 & 0.0728951 & 3.00083 & 0.215423 & 0.0601441 & 1.42447 & 3.42138 & 2 & {$(<1)$} \\
        {$\Strontium$} & {$\mathrm{S_\frac{1}{2}}$} & 2.707 & 0.0686609 & 2.9971 & 1.65612 & 0.595327 & 2.1807 & 4.11765 & 11 & 3 \\
                       & {$\mathrm{P_\frac{1}{2}}$} & 2.31 & 1.73319 & 2.95583 & 1.16722 & 0.429748 & 2.03218 & 4.00522 & 59 & 2 \\
                       & {$\mathrm{P_\frac{3}{2}}$} & 2.31 & 2.01096 & 2.96084 & 0.878713 & 0.308999 & 1.937 & 3.91793 & 34 & 4 \\
                       & {$\mathrm{D_\frac{3}{2}}$} & 1.456 & 0.057733 & 2.99692 & 0.916547 & 0.344931 & 1.90863 & 3.84306 & 7 & 2 \\
                       & {$\mathrm{D_\frac{5}{2}}$} & 1.456 & 0.0596584 & 2.99682 & 0.854725 & 0.32005 & 1.88614 & 3.82101 & 7 & 2 \\
        {$\Rubidium$}  & {$\mathrm{S_\frac{1}{2}}$} & 3.1311804 & 1.26667 & 3.0006 & 0.0371408 & 0.0422265 & 1.62668 & 3.61202 & 2 & 1 \\
                       & {$\mathrm{P_\frac{1}{2}}$} & 2.6548849 & 2.70159 & 2.99449 & 0.0333433 & 0.0414708 & 1.45403 & 3.42116 & 17 & 2 \\
                       & {$\mathrm{P_\frac{3}{2}}$} & 2.6416737 & 2.51389 & 2.99482 & 0.0323853 & 0.0402913 & 1.43772 & 3.40454 & 15 & 1 \\
                       & {$\mathrm{D_\frac{3}{2}}$} & 1.34809171 & 1.15449 & 2.98878 & 0.213663 & 0.312492 & 2.06666 & 4.01305 & 32 & 3 \\
                       & {$\mathrm{D_\frac{5}{2}}$} & 1.34646572 & 1.13923 & 2.98858 & 0.215494 & 0.317436 & 2.06616 & 4.01158 & 32 & 3 \\
        \bottomrule
    \end{tabular}
\end{table}

\end{document}